\documentclass{aastex}
\usepackage{amssymb}
\usepackage{epsfig}
\usepackage{graphicx}
\usepackage{longtable}
\usepackage{caption}
\usepackage{amsfonts}
\usepackage{amscd}
\usepackage{graphics}
\usepackage{amsmath}
\usepackage{times}
\usepackage{color}
\usepackage{footnote}
\usepackage{ulem}
\usepackage[T1]{fontenc}

\shorttitle{EBL estimation using HBL}
\shortauthors{Sinha, Sahayanathan, Misra, Godambe, Acharya}

\begin{document}
\title{Estimation of the Extragalactic Background Light using TeV Observations of BL~Lacs}
\author{Atreyee Sinha\altaffilmark{1}, S. Sahayanathan\altaffilmark{2}, R. Misra\altaffilmark{3}, S. Godambe\altaffilmark{2},
B. S. Acharya\altaffilmark{1}}

\altaffiltext{1}{Department of High Energy Physics, Tata Institute of Fundamental Research, Homi Bhabha Road, 
	Mumbai 400005, INDIA;\{atreyee,acharya\}@tifr.res.in}
	\altaffiltext{2}{Astrophysical Sciences Division, Bhabha Atomic Research Center, Mumbai, India; \{sunder,gsagar\}@barc.ernet.in}
\altaffiltext{3}{Inter-University Center for Astronomy and Astrophysics, Post Bag 4,
Ganeshkhind, Pune-411007, India; rmisra@iucaa.ernet.in}

\begin{abstract}
The very high energy (VHE) gamma ray spectral index of high energy peaked blazars correlates strongly with its corresponding redshift whereas no such correlation is observed in the X-ray or the GeV bands. 
We attribute this correlation to a result of photon-photon absorption of TeV photons with the extragalactic background
light (EBL) and utilizing this, we compute the allowed flux range for the EBL, which is independent of previous estimates. 
The observed VHE spectrum of the sources in our sample can be well
approximated by a power-law, and if the de-absorbed spectrum is also assumed to be a power law, then we show that the 
spectral shape  of EBL will be $\epsilon n(\epsilon) \sim k log(\frac{\epsilon}{\epsilon_p}) $.
We estimate the range of values for the parameters defining the EBL spectrum, $k$ and $\epsilon_p$, 
such that the correlation of the
intrinsic VHE spectrum with redshift is nullified. 
The estimated EBL depends only on the observed correlation and the assumption of a power law
source spectrum.
Specifically, it does not depend on the spectral modeling or radiative mechanism of
the sources, nor does it depend on any theoretical shape of the EBL spectrum obtained through cosmological calculations. 
The estimated EBL spectrum is consistent with the upper and 
lower limits imposed by different observations. Moreover, it also agrees closely with the 
theoretical estimates obtained through cosmological evolution models.
\end{abstract}



\section{Introduction}\label{sect:intro}

The extragalactic background light (EBL) is an isotropic diffuse radiation field
extending from Ultraviolet (UV) to Infrared (IR) wavelength ($\lambda = 0.1\,-\, 1000 \mu m$). 
It is the relic radiation containing information about the
structure formation epoch of the universe and hence, is an important
cosmological quantity \citep{dwek-review, de-ang}.  The main contributors of the EBL spectrum are
the stellar emission (peaking at optical-UV) and the dust emission (peaking at
IR). Direct measurement of EBL is very difficult
due to strong foreground contamination by the Galactic and zodiacal
light, and depends on the choice of the zodiacal light models
\citep{zl1,zl2}.
However, different upper and lower limits on EBL, based on various observations and deep galaxy number 
counts, have been put forth \citep{hauser,2000MNRAS.312L...9M,domin,2012ApJ...758L..13H}.
Theoretical prediction of the spectral
energy distribution (SED) of EBL 
can be obtained 
by evolving stellar 
populations and galaxies under various cosmological initial conditions \citep{primack, SMS,frans,gil,finke,kneiske}.
However, such models involve a large number of parameters and
the estimated EBL spectrum depends upon the
underlying assumptions (Figure \ref{fig:our_mod}) \citep{dwekold, deAng-axion, dwek-review}.

Alternatively, indirect estimation of EBL intensity can be obtained by studying 
the very high energy (VHE) gamma ray (E $> 100GeV$) spectrum of distant blazars, a class of Active Galactic 
Nuclei for which the relativistic jet is aligned close to 
the line of sight of the observer \citep{UrryPadovani95}. The VHE photons emitted from 
blazars are absorbed en-route by forming electron-positron pairs on 
interaction with the EBL photons, thereby causing the observed spectrum to differ 
significantly from the intrinsic one. The EBL intensity can thus be estimated 
from gamma ray observations of blazars under various assumptions of the intrinsic spectrum
\citep{madau, coppi}. Assuming that the EBL spectral shape is described by 
the theoretical estimates, \cite{stanev} used the VHE spectrum of Mkn501
during a flare to constrain the overall EBL intensity and corresponding spectral index.
 \cite{aha-upp} set an upper limit on the EBL intensity by assuming
that the intrinsic VHE spectral index of blazars cannot be harder than 1.5. Similar
upper limits on EBL have been put forward by various authors, based on allowed hardness 
of the intrinsic VHE spectrum \citep{guy,mazin,orr}.

The EBL estimated from the VHE spectra of blazars often depends heavily on the underlying blazar 
emission models, though their broadband emission is 
still not well understood. The SED of blazars 
is dominated by a non-thermal spectrum extending from radio-to-gamma rays and
are characterized by two peaks, with the low energy spectrum
peaking at IR--X-ray and the high energy spectrum peaking at gamma rays \citep{ghisellini2011}.
They are further classified as BL Lac objects and flat spectrum radio quasars (FSRQ), 
where FSRQ show strong emission and/or absorption lines, while such features are absent/weak in the former. 
Depending upon the location of the low energy peak, BL Lacs are further subdivided 
into low energy peaked BL Lacs (LBL), intermediate energy peaked BL Lacs (IBL) and 
high energy peaked BL Lacs (HBL) \citep{fossati,ghisellini2011}. 
The low energy emission from BL Lacs is generally interpreted as synchrotron emission from a non-thermal
distribution of electrons losing their energy in a magnetic field, while the high energy emission mechanism is still under debate.
Under leptonic origin, this emission is modeled as 
inverse Compton scattering of soft target photons by the same electron distribution
responsible for the low energy emission, whereas in hadronic models, the high energy emission 
is an outcome of hadronic processes from a region with energetic protons. 
The constraints available through present observations are not sufficient enough to differentiate 
between these models satisfactorily \citep{bottcher02}.

Before 2000, the number of blazars detected at VHE energies were few($\sim 4$), primarily 
due to low sensitivity of first generation atmospheric Cherenkov telescopes
\citep{costamante}. However, with the advent of new generation high
sensitivity telescopes, namely VERITAS, 
MAGIC and HESS, the number of blazars 
detected at this energy are more than $50$\footnote{www.tevcat.uchicago.edu}. Hence the present period allows 
one to perform a statistical study of VHE blazars to estimate the EBL, independent of various emission models. 
A study of similar kind has been
 performed by \cite{fermi-ebl} using blazars detected by the {\it Fermi}-LAT, a satellite 
based gamma ray experiment. They used the GeV spectrum of $\sim 150$ blazars to estimate 
the EBL at UV--optical wavelengths. 

In this work, we utilize a novel method to estimate the EBL spectrum at IR energies
from the observed VHE spectrum of HBL. First, we show that the observed VHE spectral 
index of HBL correlates well with the redshift. 
We attribute this correlation to a result of EBL absorption, since such correlations are 
not seen in other wavebands. 
The observed spectrum of all the sources in our sample can be well described by a power law.
Considering the source spectrum also as a power law, we show that this is expected for a 
particular shape of EBL \citep{SS06}. 
The parameters defining the EBL spectrum are then constrained by
nullifying the correlations of the intrinsic VHE spectral index with redshift.

In the next section, we present our correlation study between the observed spectral indices
and redshift to show the presence of EBL induced absorption on VHE spectra of blazars. In \S \ref{sect:Methodology},
we describe the formalism used to estimate the EBL using the correlation study and in \S \ref{sect:discussion},
we discuss the implications of the results.
A cosmology with $\Omega_M = 0.3$, $\Omega_\Lambda = 0.7$ 
and $H_0 = 71\, km\,s^{-1}\,M pc^{-1}$ is used in this work.

\section{EBL signature on VHE Spectra}\label{sect:sel}

The effect of the absorption of VHE photons by the EBL is to steepen the
VHE spectra, hence providing a signature of the EBL \citep{vassiliev,nijil}.
Since sources at higher redshifts are more affected by absorption; their
average spectra are expected to be steeper than the lower redshift ones. 
To investigate this, we perform a correlation study between the VHE spectral 
index\footnote{We define the spectral index, $\Gamma$, such that $dN/dE \propto E^{-\Gamma}$ [ $ph/cm^2/sec/TeV$ ]} 
$\Gamma$, and redshift for a homogeneous set of sources which are detected at VHE. 
We select all HBL detected by the HESS, MAGIC and VERITAS telescopes with 
known redshifts and measured spectral index. We restrict our sample to only HBL since an 
intrinsic systematic hardening with source type, from FSRQ to HBL, has been observed at 
the GeV energies \citep{fermicat}; moreover, a non-homogeneous sample may lead
to spurious correlations. This restricts the farthest source in the sample 
to be 1ES 0414+009 at a redshift of z = 0.287. In Table \ref{Table:src_list}, we list 
all the HBL detected at VHE along with ones for which the redshift information is uncertain (lower group).
Again from the list, we group 8 HBL (middle group), due to their unusual properties.
The de-absorbed VHE spectral index of these sources, obtained considering various EBL models, suggests 
their spectrum is extremely hard with index $ < 2 $ \citep{tav_ehbl,tanaka_ehbl}. Moreover, these sources are less luminous 
compared to other HBL with their synchrotron spectrum peaking at energies $ > 10 keV $.
Due to these peculiar properties, these sources have been classified as extreme HBL (EHBL) and occupy
a distinct position in the so called blazar sequence \citep{costamante}.

\begin{figure}
 \centering
 \includegraphics[width=1\linewidth]{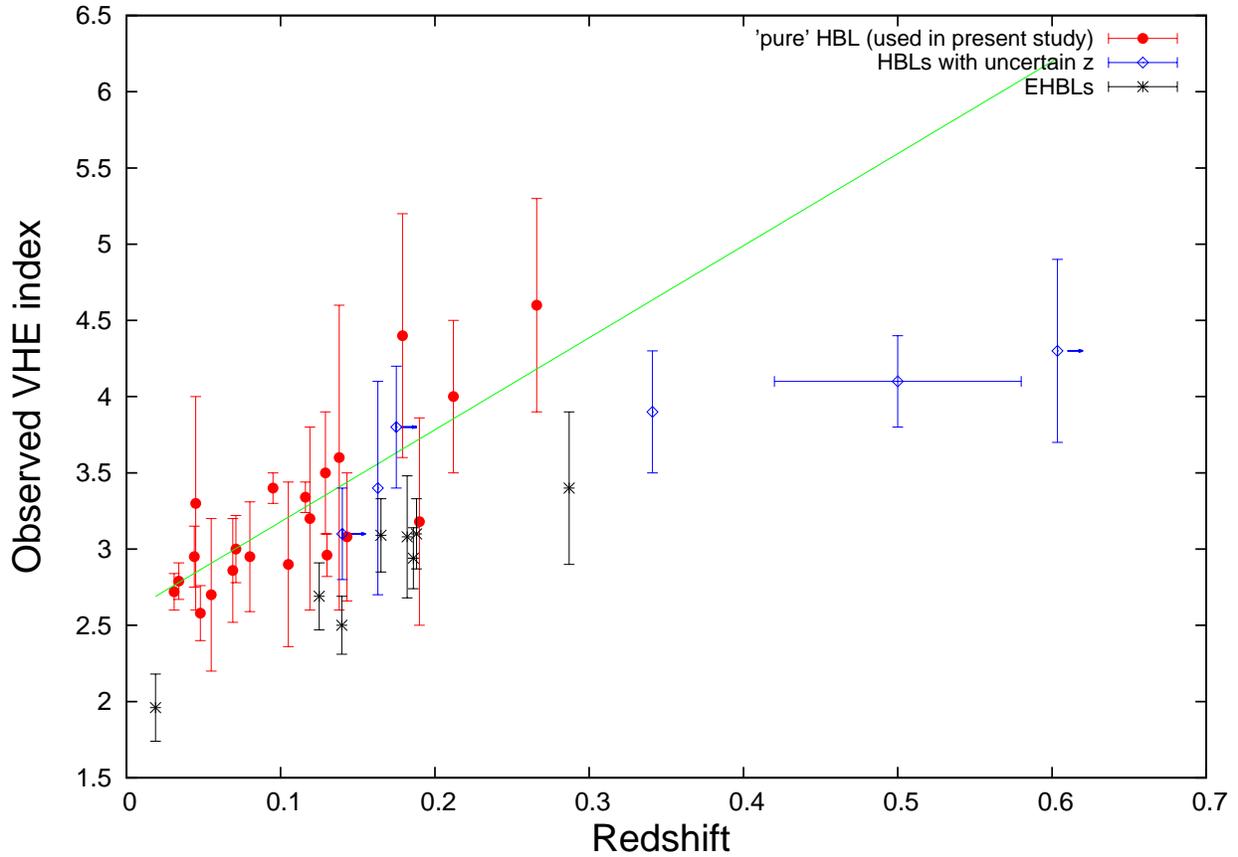}
 \caption{Distribution of the observed VHE spectral index of the selected HBL with redshift. 
 The black stars correspond to extreme HBL and blue open diamonds are the ones with 
 uncertain redshifts. The lower limits on the redshifts have been shown with solid (blue) 
 arrows. The solid line (green) is the best fit straight line to the HBL denoted by
 filled circles (red).}
 \label{fig:spec_index}
\end{figure}

In Figure \ref{fig:spec_index}, we show the variation of $\Gamma$ with redshift for all the sources 
listed in Table \ref{Table:src_list}. A Spearman rank correlation analysis on all these sources,
with known redshift, shows that they are well correlated with a rank correlation coefficient, $rs = 0.58$,
corresponding to a null hypothesis probability of $P_{rs} = 9.4 \times 10^{-4}$. Repeating the study with
EHBL removed from the list improves the correlation considerably, with $rs= 0.75$, corresponding to 
$P_{rs} = 8.02 \times 10^{-5}$.
Hence this study again suggests that, probably, EHBL can be treated as a 
separate class of HBL. However, poor statistics does not let one to assert this inference strongly.

\begin{figure}
 \centering
 \includegraphics[width=1\linewidth]{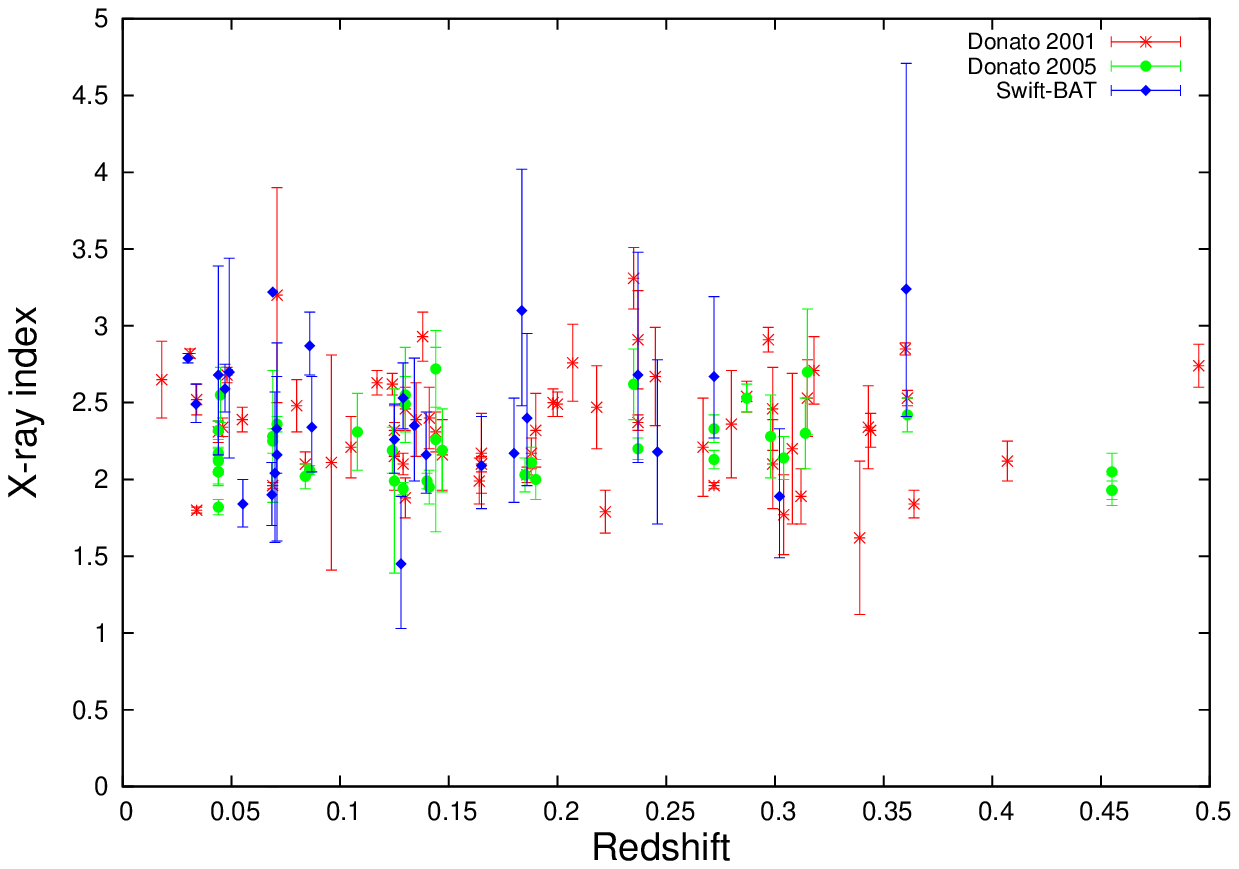}
 \caption{ Distribution of the observed X-ray spectral index of HBL with redshift. 
	 The blue diamonds are from the {\it Swift}-BAT catalog \citep{batcat}, the green circles from the {\it Beppo}-SAX catalog \citep{donato2005}, and the red stars from archival X-ray catalog  \citep{donatocat}}
 \label{fig:xrayind}
\end{figure}

Although the redshift range of the sample is small, a positive correlation may also occur due to rapid 
redshift evolution of HBL, such that the intrinsic VHE spectral index increases with redshift.  
If so, then the redshift evolution should 
be expected to have an effect on the spectral shape at other wavelengths. To examine
this possibility, we further studied the correlation between X-ray spectral indices with
redshift for HBL \footnote{In this work, we restrict our sample of HBL to only those with z<0.5, consistent with our VHE-HBL sample} using \\ a) 70 months of {\it Swift}-BAT catalog consisting of 27 HBL
\citep{batcat}, \\ b) six year {\it Beppo}SAX catalog consisting of 39 HBL \citep{donato2005}
and \\ c) archival X-ray catalog from ASCA, EXOSAT, {\it Beppo}SAX, ROSAT and EINSTEIN 
consisting of 61 HBL \citep{donatocat}. \\ We found no evidence of any correlation of the X-ray spectral index with redshift and obtained the rank correlation coefficient and null
hypothesis probability for the chosen set of catalogs as, a) $rs = 0.05$ and $P_{rs} = 0.79$ 
({\it Swift}-BAT), b) $rs = -0.07$ and $P_{rs} = 0.67$ ({\it Beppo}SAX) and c) $rs = 0.03$ 
and $P_{rs} = 0.7$ (archival).  The plot of X-ray index vs. redshift for these three catalogs is given in Figure \ref{fig:xrayind}. Spearman rank correlation study was also performed between 
low energy gamma ray (GeV) spectral index and redshift for the 62 HBL listed in the second 
catalog of {\it Fermi}-LAT \citep{fermicat}. For this, we obtained 
$rs = 0.02$ with $P_{rs}=0.85$, suggesting that these quantities are
uncorrelated. Hence, these studies violate the conjecture on redshift evolution of the spectral index of
HBL, and instead support the steepening of VHE spectral index as a result of EBL absorption.

\begin{figure}
 \centering
 \includegraphics[width=1\linewidth]{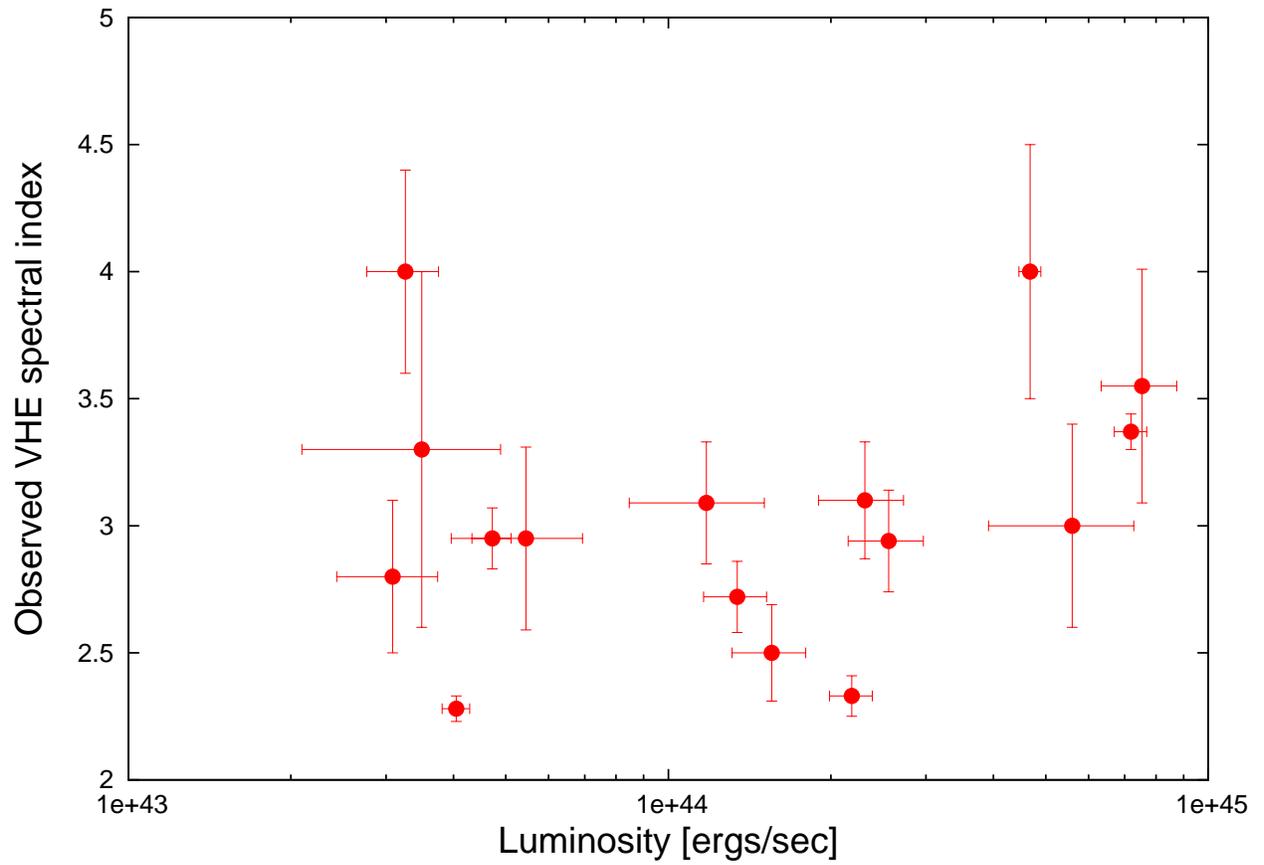}
 \caption{ Distribution of the observed VHE spectral index with luminosity of the HBL 
	 observed by the MAGIC telescope \citet{deAng-axion}.}
 \label{fig:speclum}
\end{figure}

The observed correlation between the VHE spectral index and redshift could be due to selection
effects. The luminosity is expected to correlate with redshift due to  Malmquist bias,
and if the index correlates with VHE luminosity then a correlation with redshift may
occur. However, for the HBL observed by MAGIC \citep{deAng-axion}, while the VHE luminosity does correlate with
redshift as expected, there is no significant correlation between the VHE index and luminosity. 
Here we restricted our sample only to MAGIC detected HBL as the threshold energy is different 
for each experiment.
This correlation is shown in Figure \ref{fig:speclum} and a Spearman rank analysis gives
$rs = 0.26$ and $P_{rs} = 0.34$. At X-ray energies, the correlation study between the spectral
index and X-ray luminosity resulted in a) $rs= -0.07$ and $P_{rs} = 0.15$ ({\it Swift}-BAT catalog) 
b) $rs= -0.20$ and $P_{rs} = 0.16$ ({\it Beppo}SAX catalog) and
c) $rs=-0.14$ and $P_{rs} = 0.28$ (archival). Similarly, 
correlation study between GeV spectral index and luminosity for the HBL from {\it Fermi}-LAT
catalog gives $rs= 0.11$ and $P_{rs} = 0.37$. Based on this study, we can exclude the possibility of selection effect in the observed correlation between the VHE spectral index and redshift.
Significant correlation between the difference of the VHE index and the one measured by {\it Fermi}-LAT 
with redshift was reported by several authors \citep{ss10, pradini, sanchez}. For the HBL listed in 
Table \ref{Table:src_list} (top group), we also observed significant correlation between these quantities with 
$rs = 0.71 $ and $P_{rs} = 2.2 \times 10^{-4}$; however, this correlation is weaker than the
one between VHE spectral index and redshift.

Based on these studies, it is quite evident that the
correlation between VHE spectral index and redshift can be 
attributed solely due to the effect of EBL induced 
absorption and that the intrinsic spectral index is uncorrelated with redshift.

\section{EBL Estimation}\label{sect:Methodology}
The observed VHE spectra of the HBL are well reproduced by a power law, and hence
the observed flux, $F_o(E_i)$, from a source at redshift $z$ will be 
\begin{align}
	\label{eq:obsflx}
	F_o(E_i) &= F_i(E_i)e^{-\tau(E_i,z)}\\
	 &\propto E_i^{-\Gamma} \nonumber 
\end{align}
where $F_i(E_i)$ is the de-absorbed flux at energy $E_i$ and $\tau$, the optical depth due to EBL absorption given by \citep{gould}
\begin{align}
\label{eq:tau}
\tau(E_i, z) = \int\limits_0^{z} { d}z^\prime \frac{{ d}l}{{ d}z^\prime}
\int\limits_{-1}^{1} { d\mu}\frac{(1-\mu)}{2}
\int\limits_{\epsilon_{th}}^\infty { d} \epsilon_{z^\prime} n(\epsilon_{z^\prime}, z^\prime) 
\sigma_{\gamma \gamma} ( E_i, \epsilon_{z^\prime}, \mu) 
\end{align}
Here, 
\begin{equation}
\label{lungh}
\frac{dl}{dz^\prime} = \frac{c}{H_0} \frac{1}{(1 + z^\prime) \sqrt{{\Omega}_{\Lambda} + {\Omega}_M (1 + z^\prime)^3}},
\end{equation}
is the distance traveled by a photon per unit redshift with $c$ as the velocity of light,
$n$ is the number density of the EBL photon of energy 
$\epsilon_{z^\prime}$[$=\epsilon_0(1+z^\prime)$] at redshift 
$z^\prime$, corresponding to a photon energy $\epsilon_0$
at $z=0$, $\epsilon_{th}$ [$ = 2m_e^2c^4\;(1+z^\prime)/(E_i(1-\mu))$]  is the threshold soft photon energy with $\mu$ 
the cosine of the interaction angle and the pair production cross section, $\sigma_{\gamma \gamma}$, is given by
\begin{equation}
\sigma_{\gamma \gamma}(E,\epsilon,\mu) = \frac{2\pi\alpha^2}{3m_e^2} (1-\beta^2)\times \left[2 \beta( \beta^2 -2)+(3 - \beta^4)\,ln \left(\frac{1+\beta}{1-\beta}\right)\right]
\end{equation}
with
$\beta(E,\epsilon)=\sqrt{1 - (2\;m_e^2\;c^4)/(\epsilon\,E\,(1-\mu))}$
being the speed of the electron/positron in the centre of mass frame, $\alpha$
is the fine structure constant
and $m_e$ is the electron rest mass.
Since VHE sources are detected only at 
low redshifts, one can neglect the evolution of EBL and hence \citep{madau},
\begin{equation}
	\label{eblev}
	n(\epsilon_z, z) \approx (1+z)^3 n(\epsilon_z)
\end{equation}
For an isotropic EBL distribution, the angle integrated cross section \citep{gould1404, brown}
\begin{equation}
	\bar{ \sigma}_{\gamma \gamma}(E\epsilon) = \frac{1}{2}\int\limits_{-1}^{1} { d\mu} (1 - \mu) \sigma_{\gamma \gamma}(E,\epsilon,\mu)
	\label{eq:sigmabar}
\end{equation}
peaks at $E\epsilon = 3.56\; m_e^2\;c^4$. Approximating $\bar{\sigma}_{\gamma\gamma}$ as a delta function along with $E_z \approx E_i$ and 
$\epsilon_z\approx \epsilon_0$, equation (\ref{eq:tau}) can be simplified to
\begin{align}
	\tau(E, z) \approx  A_{\gamma \gamma}\,\epsilon \,n(\epsilon) f(z)
   \label{eq:tau_delf}
\end{align}
where $A_{\gamma \gamma}$($\approx 3.7 \times 10^{-26}\, cm^2$) is a constant, $\epsilon \approx  \frac{3.6\, m_e^2 c^4}{E}$ and $f(z)$ is given by
\begin{align}
f(z) &= \int\limits_0^{z} { d}z^\prime \frac{{ d}l}{{ d}z^\prime} (1+z^\prime)^3 \\
&\approx \frac{c}{H_0}z \nonumber
\end{align}
If the source spectrum is assumed to be a power-law, $F_i(E) \propto E^{-\zeta}$, then using equations 
(\ref{eq:obsflx}) and (\ref{eq:tau_delf}) we obtain
\begin{align}
 \label{eq:ebl_local}
	\epsilon \, n(\epsilon) = k \, ln\left(\frac{ \, \epsilon_p}{\epsilon}\right)
\end{align}
where  $k = \frac {\Gamma-\zeta}{A_{\gamma \gamma}f(z)}$ and 
$\epsilon_p \approx \frac{3.56 \; m_e^2c^4}{E^\star}$ are source independent constants with 
$E^\star$ being the energy of the gamma ray photon at which the EBL induced absorption is negligible.
While this form of the EBL spectrum is derived for the approximate optical depth equation (\ref{eq:tau_delf}), we have verified numerically that for the
range of redshifts considered here, this EBL spectral shape will result in a nearly power-law observed spectrum. In particular, we have verified that if the EBL spectra is defined by equation (\ref{eq:ebl_local}) and the absorption optical depth is given by equation (\ref{eq:tau}), the observed VHE spectrum for a source at $z=0.3$ can be well described as a power-law. The deviation from a power-law for sources with $z < 0.3$ is less than 10 \%.
Hence for all calculations in this work we have used equation (\ref{eq:tau}) for the optical depth.
It may be noted that \cite{SS06} obtained a similar form of the optical depth while approximating a theoretical EBL spectrum by an analytic expression. However, here we arrive at this form of the EBL spectrum equation (\ref{eq:ebl_local}) only from the criterion that both the
intrinsic and absorbed VHE spectra are well described by a power-law and
hence our approach is independent of any cosmological calculations.

The EBL spectrum given by equation (\ref{eq:ebl_local}) is characterized by two constants, 
namely $k$ and $\epsilon_p$, which in turn determines the source spectral index. Since the source 
spectral index should be uncorrelated with redshift (\S \ref{sect:sel}), the allowed range of
$k$ and $\epsilon_p$ is restricted. For $k = 2.4  \times 10^{-3} cm^{-3}$ and 
$\epsilon_p = 4.6$ eV, we found that the computed source spectral indices, for the sources listed 
in Table \ref{Table:src_list} (top group), turn out to
be ``most'' uncorrelated with $rs=0.001$ and a maximum null hypothesis probability $P_{rs} =0.99$. 
Within $1-\sigma$ confidence limit, corresponding to $P_{rs}>0.33$, we obtained the range of
$k$ and $\epsilon_p$ as $2.2^{+1.6}_{-0.7} \times 10^{-3}\, cm^{-3}$ and $4.6^{+4.4}_{-3.4}\, eV$, respectively.
The resultant EBL spectrum, consistent with these values of $k$ and $\epsilon_p$,  
is plotted in Figure \ref{fig:our_mod} along with the constraints derived from various 
observations \citep{dwek-review} and other theoretical estimates \citep{gil,frans,finke,kneiske}. If we consider all the HBL with known 
redshift in Table \ref{Table:src_list} (top and middle group), the $1-\sigma$ confidence limit of
$k$ and $\epsilon_p$ are $2.8^{+2.6}_{-1.8} \times 10^{-3}\, cm^{-3}$ and $5.2^{+4.8}_{-4.0}\, eV$, respectively.

\begin{figure}
	\includegraphics[width=1\linewidth]{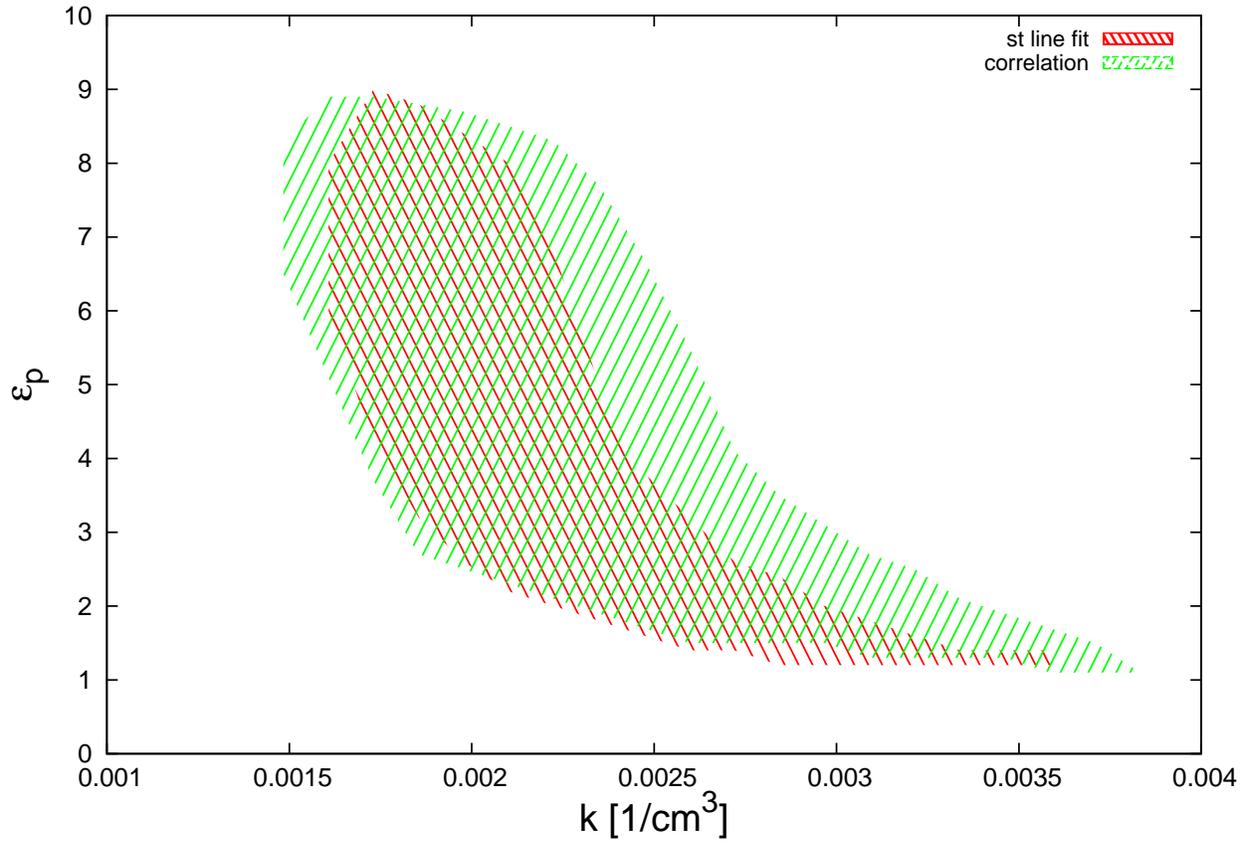}
	\caption{The $1-\sigma$ confidence region for the parameters $k$ and $\epsilon_p$ 
	obtained from the correlation study (green forward stripes) and the straight 
line fit (red backward stripes)} 
	\label{fig:ellipse}
\end{figure}

Alternatively, a linear fit over source spectral index versus redshift can be used to constrain the constants
$k$ and $\epsilon_p$. A linear fit between $\Gamma$ and redshift resulted in a straight line of
slope $6.0\pm1.1$ with reduced chi-square $\chi_{red}^2 \approx 1.1$. Since the source spectral
index is uncorrelated with the redshift (\S \ref{sect:sel}), a linear fit between these quantities should result in a 
constant line (a line with slope $0$). Within 1-$\sigma$ confidence limit, this corresponds to an
EBL spectrum with $k = 2.4^{+1.2}_{-0.8} \times 10^{-3}\,  cm^{-3} $ and $\epsilon_p = 5.2^{+3.8}_{-4.0}\, eV$, for which this condition can be 
achieved. We find that these constraints on $k$ and $\epsilon_p$ are consistent with the one obtained 
earlier through nullifying the correlation between source VHE spectral index and redshift. 
In Figure \ref{fig:ellipse}, we show the allowed range of $k$ and $\epsilon_p$ obtained by these
two methods.

\section{Discussions}\label{sect:discussion}

\begin{figure}
	\includegraphics[width=1\linewidth]{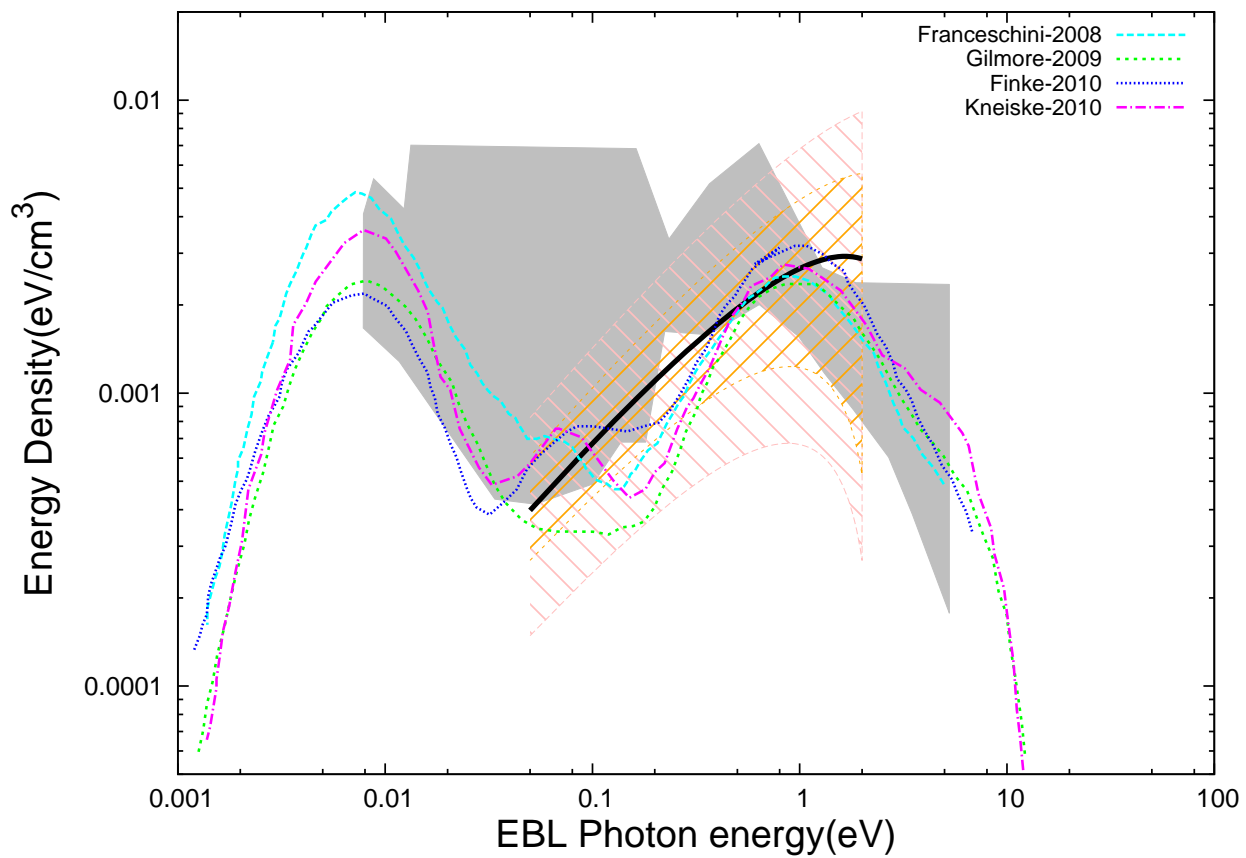}
	\caption{The best fit EBL spectrum estimated in this work (thick black line) and the $1-\sigma$ (checkered orange region) and $2-\sigma$ constrains (striped pink region) 
compared with the different 
theoretical models of Franceschini (\cite{frans}), Gilmore (\cite{gil}), Finke (\cite{finke}) and Kneiske (\cite{kneiske}) .
The solid grey region shows the upper and lower limits estimated from various observations \citep{dwek-review}.}
	\label{fig:our_mod}
\end{figure}

The EBL spectrum presented in this work is estimated directly from the
observed VHE spectra of HBL with the condition that the source spectrum
should be uncorrelated with redshift. The main uncertainty lies in the
assumption that the source VHE spectrum is a power law, and that approximate
EBL spectral shape is given by equation (\ref{eq:ebl_local}). However, the 
latter is verified numerically to reproduce
the observed spectrum which can be well represented by a power law.
Interestingly, the present estimation does not depend on the nature of the 
radiative process active in HBL or dust/stellar emission models from galaxies,
yet still agrees well with other estimates as shown in Figure \ref{fig:our_mod}. 
Moreover, though the 1-$\sigma$ uncertainty range on the EBL spectrum is 
nearly a factor $\sim 4$, it is competitive compared to constraints put by 
observations and by other estimates. The predicted EBL spectrum 
is reasonably confined within the upper and lower limits (grey shaded area 
in Figure \ref{fig:our_mod}), obtained independently through 
observations (see \S 1). When
compared with the other EBL estimates, obtained through cosmological 
evolution models, the present one predicts stronger
emission at lower energies but closely agrees at higher energies,
though the predicted spectrum is not well constrained in this regime.

Deviation of the source spectrum from a power law may 
modify the EBL spectral shape described by equation 
(\ref{eq:ebl_local}) considerably.  In such case, the present formalism needs 
to be modified by studying the correlation of other suitable observables instead of 
the power law spectral indices. However, the source spectrum of 
HBL (z<0.3) obtained using various EBL models are fitted resonably well by a 
power law and the one presented here agrees closely with these EBL models. To investigate  
further, we repeat the study considering the EBL spectral
shape due to \cite{frans}, \cite{gil}, \cite{finke} and \cite{kneiske}. 
Following \cite{abdoebl, fermi-ebl, hessebl, rebecca}, we define the observed
spectrum to be
\begin{equation}
	F_o(E) = F_i(E_z)exp(-\tau_{theory}(E_z,z)\times b)
	\label{compmod}
\end{equation}
where $\tau_{theory}$ is the optical depth predicted by the above mentioned theoretical 
models, and $b$ is a normalization factor required to assure that the de-absorbed spectral
index
is uncorrelated with the redshift. Then, $b = 1.0 $ would imply that 
the particular model is consistent with the non correlation of the source VHE index
with redshift. For the models discussed above, we obtained 
$b_{Kneisnke} = 1.5^{-0.6}_{+0.7}$, $b_{Franschini} = 1.4^{-0.6}_{+0.7}$, 
$b_{Finke} = 1.1^{-0.5}_{+0.6}$ and $b_{Gilmore} = 1.6^{-0.5}_{+0.7}$.
From these, one can argue that the EBL model due to  \cite{finke} 
 very well supports the non 
correlation of the source VHE index with redshift; whereas, the deviation
from this condition is observed to be maximum in case of \cite{gil}.

\begin{figure}
	\includegraphics[width=1\linewidth]{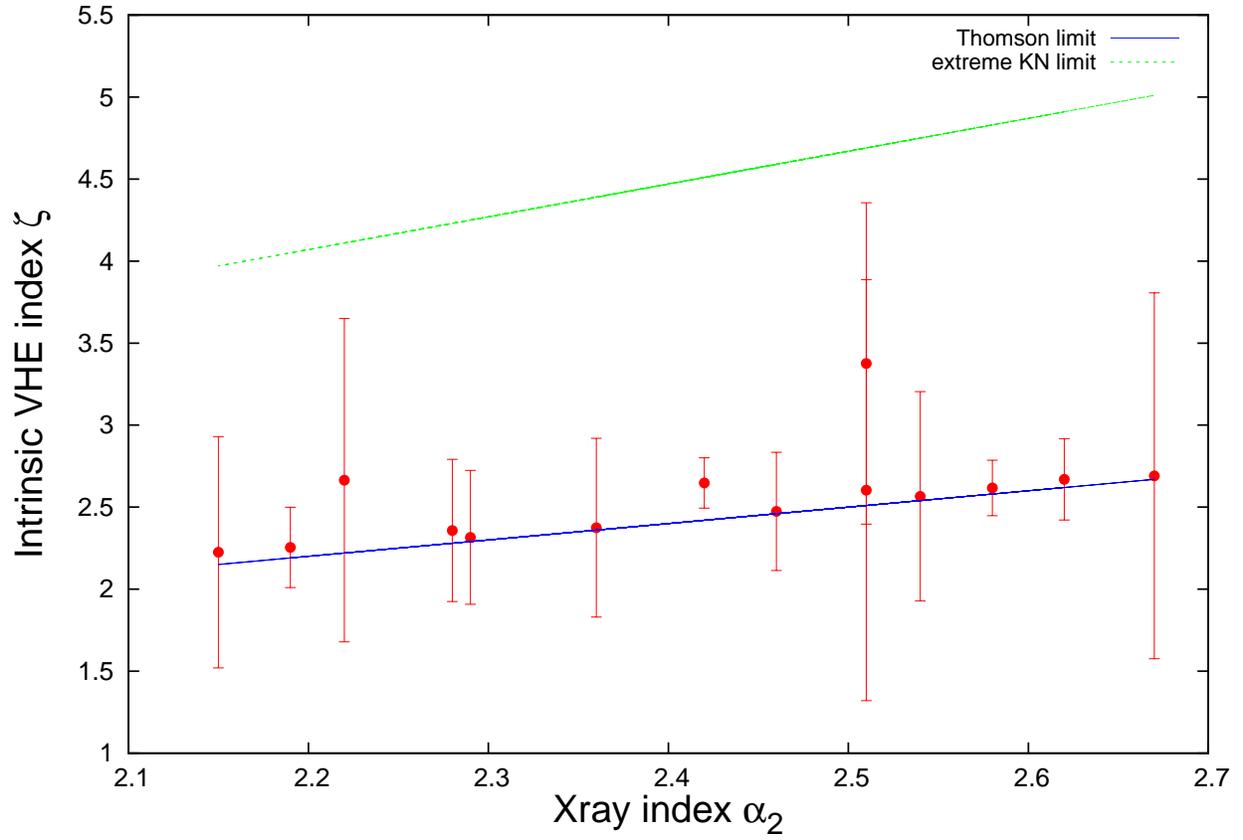}
  	\caption{Distribution of the calculated intrinsic TeV indices with X-ray indices. 
		Solid line (blue) denotes the Thomson regime, whereas the dashed line (green)
	denotes extreme Klein-Nishina regime.}
  	\label{fig:sanity}
\end{figure}

The EBL spectrum, estimated in this work, can be used to find the 
intrinsic spectral index of HBL, which can then be compared with the one
predicted by the radiative models of HBL.
Under leptonic models, the spectral energy distributions of HBL are well reproduced by considering
synchrotron and synchrotron self Compton emission from a
broken power-law distribution of electrons. In such a case, the VHE spectrum
corresponds to the high energy tail of the electron distribution. 
Similarly, the X-ray spectrum lying beyond the synchrotron peak is also  
governed by the high energy end of the electron distribution. Indeed, the
X-ray-TeV correlation observed during flares further suggests that
the same electron distribution is responsible for the emission at these 
energies \citep{xraytev}. Hence, it can be argued that the spectral index at 
these energies is related to the high energy particle spectral index 
of the underlying electron distribution. If the Compton scattering 
responsible for the VHE emission occurs in the Thomson regime, then the 
corresponding spectral index $\zeta$ will be same as the X-ray spectral
index, $\alpha_2$. On the other hand, if the scattering process happens at the
extreme Klein-Nishina regime, then the VHE spectral index will be $2\alpha_2-\alpha_1$ \citep{tavKN},
where $\alpha_1$ is the optical spectral index reflecting the low
energy electron spectral index. In general, the VHE index is expected to lie in between
these two limits. To examine this, we compare the 
intrinsic VHE spectral index, computed in this work, with 
the X-ray spectral indices 
of the sources for which
simultaneous/contemporaneous observations are available
from {\it Swift}-XRT/{\it Suzaku}/{\it XMM}-Newton/{\it Swift}-BAT observations (Table  \ref{Table:src_list}).
In Figure \ref{fig:sanity},
we plot the intrinsic VHE spectral indices against the X-ray 
spectral indices with the limiting lines corresponding to
Thomson and extreme Klein-Nishina regimes. For the latter limit,
we assume the optical spectral index as $1/3$ since this limits the 
hardest synchrotron spectrum attainable \citep{pacholczyk}.
Interestingly, all the sources are constrained well within these
limits thereby supporting the afore mentioned interpretation.
	
The analysis done in this work was possible because the intrinsic variation of
spectral index for HBL is relatively small.  The fractional root mean square deviation ($f_{rms}$)
of the de-absorbed VHE indices is $0.16$, which is comparable to that of the X-ray ($f_{rms,X}=0.14$) and
the low energy GeV gamma-rays ($f_{rms,GeV}=0.12$). 
The VHE index  $f_{rms}$ is significantly
less than the index change due to absorption  $\Delta \Gamma \sim 2$ 
at a redshift of $z = 0.266$.
If the variation of index was comparable to the change due to absorption, 
the effect would not have been
detectable. Since $f_{rms}$ is considerably smaller than $\Delta \Gamma$, this leaves the
exciting possibility that the uncertainty in EBL, predicted by the present study, can be 
significantly reduced  with increased number of blazars detected at VHE energies.

The work presented here is similar to the EBL upper limits proposed by 
\cite{schroedter} and \cite{finkerazzaque}. They studied the steepening
of VHE spectral index with increase in redshift and attributed it to the
absorption by EBL. \cite{schroedter} suggested an upper limit on EBL 
by assuming that the source spectral index, $\zeta$, cannot be harder 
than $1.8$, whereas \cite{finkerazzaque} considered limits for 
$\zeta$ as $1$ and $1.5$. Following a similar procedure, \cite{yangwang}
proposed an EBL upper limit by considering {\it Fermi}-LAT spectral index as
an allowed limit on $\zeta$. In this work, we systematically study
the steepening of the VHE spectra of HBL with respect to redshift and
exploit it to estimate the EBL spectrum. In addition, we do not impose
any limits on $\zeta$; instead, the derived value of $\zeta$, using the 
current EBL,
lies well within our present understanding of blazar emission models
(Figure \ref{fig:sanity}).

Lately, various EBL estimates have been proposed, exploiting the properties of 
TeV blazars under unique techniques. \cite{mazin} and \cite{meyer} 
estimated an upper limit on EBL
by employing splines. From the observed VHE
spectra of blazars, they converged to a particular shape of EBL, which 
leads to a de-absorbed spectra that are physically acceptable under 
the present understanding of blazars. \cite{nijil, dom2013}
and \cite{sunder_ebl} reproduced the broadband SED of VHE blazars under
leptonic model and thereby predicting the intrinsic VHE spectra. Comparing
this with the observed VHE spectra, they estimated the optical depth for
the attenuation of VHE gamma rays. While \cite{nijil} used this to show the
inconsistency among various EBL models interpreted theoretically,
\cite{sunder_ebl} showed a systematic deviation of the optical
depths towards high energy; between the estimated and the ones 
predicted by various EBL models. \cite{dom2013} used the 
estimated optical depths to determine the cosmic gamma ray horizon.
\cite{rebecca} considered the EBL models by \cite{frans,gil,domin} to estimate the
optical depth for the VHE sources at $z\sim 0.1$. The optical depth is
then scaled by a parameter to reproduce the observed flux for 
a range of de-absorbed spectral indices. Based on this scaling 
parameter, they concluded that these EBL models are consistent 
with the observed spectra, though the error on the parameter is
large. Unlike these models, the work presented here does not have any bias
on blazar emission models or a particular
EBL shape. Instead, it relies upon the observed correlation between
the VHE spectral index and redshift, along with the assumption that the de-absorbed
VHE spectra is a power law.

Finally, like other EBL models, the EBL spectrum 
presented in this work
predicts very large opacities for VHE photons from distant sources, 
e.g. 3C\,279 at $z=0.536$ and PKS\,1424+240 at $z > 0.6$
\citep{opacity,furniss}. This is evident from Figure \ref{fig:spec_index}, 
where deviation of the observed VHE index from the best fit line is 
large for distant sources. 
This remains an open problem and may possibly be related
to VHE emission through secondary processes resulting from the development of electromagnetic  and hadronic cascades
in the intergalactic medium \citep{cascade} or more exotic scenarios associated with creation of axion like particles \citep{deAng-axion}.
With the help of
present high sensitivity VHE telescopes and future telescopes, 
like CTA (Cherenkov Telescope Array),
these uncertainties can be cleared, providing more insight into our 
cosmic evolution.

\section{Acknowledgements}

The authors thank the anonymous referee for his valuable comments and suggestions. AS thanks Varsha Chitnis for helpful discussions and support.
This research has made use of TeV catalog (http://tevcat.uchicago.edu/) created and maintained by
Scott Wakely And Deirdre Horan and partially supported by NASA and the NSF.

\newpage

\begin{longtable}{|l|c|c|c|c|c|c|}
 \hline
 \scriptsize
 Source name & 		$z$ &  $\Gamma$ & $E_{VHE}$ & $\alpha_2$ & $\alpha_f$ & Ref\\ \hline
 Mkn421 &		0.031 &	 2.72 $\pm$	0.12 &   0.2 -	10    & 2.58 $\pm$ 0.03& 	1.771 $\pm$	0.012& 1 \\
 Mkn501 &		0.034 &	 2.79 $\pm$	0.12 &   0.1 -	2.0   &	2.42 $\pm$ 0.01 &	1.738 	$\pm$ 0.027 & 2 \\ 
 1ES 2344$+$514$^*$ &	0.044 &	 2.95 $\pm$	0.20 &	 0.2 -  2.0   &	2.62 $\pm$ 0.50& 	1.716 $\pm$	0.08& 3 \\
 Mkn180 & 		0.045 &  3.30 $\pm$ 	0.70 &   0.2 -  6.6   & - 	&		1.74 $\pm$	0.083& 4 \\
 1ES 1959$+$650 &	0.048 &	 2.58 $\pm$	0.18 &	 0.2 -	2.5   &	2.19 $\pm$ 0.02& 	1.937 $\pm$      0.031 & 5\\
 1ES 1727$+$502 &	0.055 &	 2.70 $\pm$	0.50 &	 0.15 -  2.0  &	- 	&		2.0 $\pm$ 	0.2  & 6  \\   
 PKS 0548$-$322$^*$ &	0.069 &	 2.86 $\pm$	0.34 &	 0.3 -	4.0   &	2.28 $\pm$ 0.23& 	   -                  & 7\\
 PKS 2005$-$489 &	0.071 &	 3.00 $\pm$	0.22 &	 0.4 -	4.0   &	2.46 $\pm$ 0.01& 	1.779    $\pm$  0.047 & 8\\
 RGB J0152+017 & 	0.080 &  2.95 $\pm$ 	0.36 & 	 0.1 -  4.0   & - 	&		1.788  $\pm$    0.137 & 9\\
 BZB J0013-1854 & 	0.095 &  3.4  $\pm$ 	0.10 &   0.2 -  2.0   & - 	&		1.96 $\pm$	0.2 & 10\\
 1ES 1312-423 & 	0.105 &  2.85 $\pm$ 	0.7  &   0.2 -  4.0   & -       &               1.4 $\pm$	0.4   & 11\\
 PKS 2155$-$304 &	0.116 &	 3.34 $\pm$	0.10 &	 1.0 -	10.   & 2.36 $\pm$ 0.01& 	1.838    $\pm$  0.015 & 12 \\
 B3 2247$+$381  & 	0.119 &  3.20 $\pm$ 	0.60 &   0.1 -  2.0   & - 	&		1.837    $\pm$  0.113 & 13 \\
 H 1426$+$428$^*$ &	0.129 &	 3.50 $\pm$	0.40 &	 0.3 -	2.00  &	2.54 $\pm$ 0.24& 	1.316    $\pm$  0.123 & 14 \\
 1ES 1215$+$304  &	0.130 &	 2.96 $\pm$	0.14 &	 0.1 -	1.51  &	2.29 $\pm$ 0.16& 	2.019    $\pm$  0.036 & 15 \\
 1ES 0806$+$524 &	0.138 &	 3.60 $\pm$	1.00 &	 0.3 -	1.02  &	2.67 $\pm$ 0.08& 	1.938     $\pm$ 0.057 & 16 \\
 BZB J1010$-$3119 &	0.143 &  3.08 $\pm$	0.42 &	 0.25 - 3.0   & 2.15 $\pm$ 0.06& 	2.239    $\pm$  0.142 & 17 \\
 RX J0648$+$1516 &	0.179 &	 4.40 $\pm$	0.80 &	 0.2 -	0.65  &	2.51 $\pm$ 0.06& 	1.737    $\pm$  0.106 & 18 \\
 RBS 0413 &		0.190 &	 3.18 $\pm$	0.68 &	 0.3 -	1.0   &	2.22 $\pm$ 0.07& 	1.551    $\pm$  0.112 &  19 \\
 1ES 1011$+$496 &	0.212 &	 4.00 $\pm$	0.50 &	 0.15 - 0.8   &	- 	&		-         & 20 \\
 PKS 0301$-$243	&	0.266 &  4.60 $\pm$ 	0.70 &   0.1 - 5      & 2.51 $\pm$ 0.1&  	1.938  $\pm$    0.031 & 21 \\ \hline \hline
 IC310   	&	0.019 &	 1.96 $\pm$     0.22 &	 0.12 -  8.1  & - 		& &22 \\  
 RGB J0710$+$591$^*$ &	0.125 &	 2.69 $\pm$	0.22 &	 0.3 -	7.9   &	2.29 $\pm$ 0.26 & &23 \\
 1ES0229+200 	&	0.140 &	 2.50 $\pm$     0.19 &	 0.5 -  15.0  &	2.16 $\pm$ 0.28 & &24   \\
 H 2356-309 &		0.165 &	 3.09 $\pm$	0.24 &	 0.2 -	1.04  &	2.43 $\pm$ 0.11 & &25 \\
 1ES 1218+304	&	0.182 &  3.08 $\pm$ 	0.40 & 	 0.1 - 2      & - 		& &26 \\
 1ES 1101$-$232 &	0.186 &	 2.94 $\pm$	0.20 &	 0.1 -	0.66  &	2.32 $\pm$ 0.02 & &27 \\
 1ES 0347$-$121$^*$ &	0.188 &	 3.10 $\pm$	0.23 &	 0.3 -	3     &	2.27 $\pm$ 0.30 & &28 \\
 1ES 0414$+$009 &	0.287 &	 3.40 $\pm$	0.50 &	 0.2 -	0.70  &	2.40 $\pm$ 0.10 & &29 \\ \hline \hline
 HESS J1943+213  &	0.14  &  3.1  $\pm$     0.30 &  &  & &30 \\  
 1ES 1440+122 	& 	0.163 &  3.4  $\pm$     0.7 &  &  &  &31\\
 PKS 0447-439 	&	0.175 &	 3.8  $\pm$ 	0.4 & &  & &32\\
 1ES 0502+675 	& 	0.341 &  3.9  $\pm$     0.4 & & & &33\\
 PG 1553+113 	&	0.5 $\pm$ 0.08 &	4.1 $\pm$  0.3 & & & &34 \\
 PKS 1424+240   &       0.604  & 4.2 $\pm$ 0.7 & & & &35\\ \hline

  \caption{The list of HBL detected in VHE. The middle group lists the 
	  extreme HBLS and the bottom one with uncertain redshift. 
	  Column description 1: the Source name, 
	  2: the redshift ($z$), 3: Observed VHE spectral index ($\Gamma$), 4: Observed VHE energy range (in TeV), 
	  5: X-ray Spectral Index (quantities with * are obtained from {\it Swift}-BAT) 6. The {\it Fermi}-LAT average index, and 
 7: References: \newline 1. \cite{Mkn421}  2. \cite{Mkn501} 
 3. \cite{1es2344} 4. \cite{mkn180} 5. \cite{1es1959} 
 6. \cite{1727}    7. \cite{pks0548} 8. \cite{pks2005} 9. \cite{rgbj0152} 
 10. \cite{shbl} 11. \cite{1es1312} 12. \cite{pks2155} 13. \cite{b3} 
 14 .\cite{1426} 15.\cite{1es1215} 16.\cite{1es0806} 
 17. \cite{rxs} 18. \cite{rx} 19. \cite{rbs0413} 
 20. \cite{1es1011} 21. \cite{pks0301} 22. \cite{IC310}
 23. \cite{rgb0710} 24. \cite{0229}    25. \cite{h2356}
 26. \cite{1es1218} 27. \cite{1es1101} 28. \cite{1es0347}
 29.\cite{1es0414} 30 - 35. www.tevcat.edu
 }
 \label{Table:src_list}
 \end{longtable}


\bibliographystyle{apj}
\bibliography{eblref}

\end{document}